# Modified Szabo's wave equation models for lossy media obeying frequency power law

W. Chen[a)] (corresponding author) and S. Holm [b)]

Simula Research Laboratory

P. O. Box. 134

1325 Lysaker

Oslo, Norway

[a)]E-mail: wenc@simula.no

[b)]E-mail: sverre.holm@simula.no

**Running title**: Comments on Szabo's lossy media models

**Abstract**

Szabo's models of acoustic attenuation (Szabo 1994a) comply well with the empirical frequency power law involving non-integer and odd integer exponent coefficients while guaranteeing causality, but nevertheless encounter the troublesome issues of hyper-singular improper integral and obscurity in implementing initial conditions. The purpose of this paper is to ease or remove these drawbacks of the Szabo's models via the Caputo fractional derivative concept. The positive time fractional derivative is also first introduced to include the positivity of the attenuation possesses.

## INTRODUCTION

The effect of the attenuation plays a prominent role in many acoustic and ultrasound applications, for instance, the ultrasound second harmonic imaging and high intensity focused ultrasound beam for therapeutic surgery. Due to the fractal microstructures of media, such acoustic attenuation typically exhibits a frequency dependency characterized by a power law

$$S(\bar{x} + \Delta\bar{x}) = S(\bar{x})e^{-\alpha(\omega)\Delta\bar{x}}, \qquad (1)$$

$$\alpha(\omega) = \alpha_0|\omega|^y, \qquad y \in [0,2], \qquad (2)$$

where $S$ represents the amplitude of an acoustic field variable such as velocity and pressure; $\omega$ denotes the angular frequency; $\Delta\bar{x}$ is wave propagation distance; and the tissue-specific coefficients $\alpha_0$ and $y$ are empirically obtained by fitting measured data. The power law (2) is applicable for a broad range of frequency of practical interest. It has long been known that the time-space mathematical model of the attenuation formulas (1) and (2) with a power $y \neq 0,2$ is not easily constituted with the standard partial differential equation (PDE) methodology. Thus, the attenuations involving $y \neq 0,2$ are also called the anomalous attenuations.

Recent decades have witnessed great effort devoted to developing a variety of the anomalous attenuation models. These modeling approaches, however, have their respective advantages and disadvantages, for example, the frequency-domain model via the Laplace transform is simple but limited to linear cases and also computationally expensive (Wismer and Ludwig, 1995; Rossikhin and Shitikova, 1997; Ginter, 2000); the adaptive Rayleigh model (Wojcik et al, 1995) is as easy as the common PDE models to be implemented but may not be feasible to important broadband pulse propagation (Chen and Holm, 2002); the multiple relaxation model (Nachman et al, 1990) works well for relaxation-dominated attenuations but requires large amounts of computational effort (Yuan et al, 1999) and demands strenuous effort to estimate many obscure relaxation parameters and may not be applicable to anomalous attenuations involving singular memory processes (Hanyga, 1999); the fractional calculus models have better accuracy for general anomalous attenuations with fewer necessary parameters (Baglegy and Torvik, 1983; Makris and Constantinou, 1991; Ochmann and Makarov, 1993; Enelund, 1996; Gaul, 1999) but are mathematically more complicated and inflict nontrivial cost of the numerical solution. Recently, Szabo (1993, 1994a,b) presented the convolution integral wave models which comply well with arbitrary power law frequency dependences while guaranteeing causality and involving only the parameters $\alpha_0$ and $y$. However, it is not straightforward to implement the initial conditions in the Szabo's models. In addition, the severely hyper-singular improper integral hinders the numerical solution of the models.

The paper aims to remedy the above-mentioned drawbacks of the Szabo's models. We find that the Szabo's models can be recast with the Riemann-Lioville fractional derivative for non-integer power $y$. Then, the basic strategy of this study is that we replace the Riemann-Liouville fractional derivative with the Caputo fractional derivative to derive a modified Szabo's model, where the initial conditions can be easily prescribed and the hyper-singular integral is regularized inherently. To simplify the expression and formalize for potential applications in a standard setting, we also introduce a positive time fractional derivative to include the principle of causality in modeling of the frequency power law attenuation with arbitrary exponent $y$.

**I. SZABO'S CONVOLUTIONAL INTEGRAL MODEL**

By a thorough examining of the dispersion equations of the frequency-independent damped wave equation and frequency-squared dependent thermoviscous wave equation, Szabo (1994a) presented a general dispersion equation for arbitrary power $y$ of frequency dependent attenuation

$$k^2 = (\omega/c_0)^2 + i2(\omega/c_0)a_0|\omega|^y, \qquad (3)$$

where the wave number $k=\beta+i\alpha$ corresponds to the Fourier transform of the space variable. $\beta$ and $\alpha$ are respectively related to the dispersion and the attenuation. For more details on the induction of the dispersion equation (3) see Szabo (1994a). Under the so-called conservative smallness condition (equation 22 of Szabo, 1994a)

$$\alpha/\beta \approx \alpha_0|\omega|^y_{\lim}/\beta_0 = \alpha_0|\omega|^{y-1}_{\lim}c_0 \leq 0.1, \qquad (4)$$

the dispersion equation (3) approximates the power law (2) quite well and leads to the Szabo's time convolutional integral model

$$\Delta p = \frac{1}{c_0^2}\frac{\partial^2 p}{\partial t^2} + \frac{2\alpha_0}{c_0} L_{\alpha,y,t} * p, \qquad (5)$$

where * denotes the convolution operation, and

$$L_{\alpha,y,t} = \begin{cases} \delta(t), & y = 0, \\ \dfrac{2}{\pi t^3}, & y = 1, \\ -\dfrac{\Gamma(y+2)\cos[(y+1)\pi/2]}{\pi|t|^{y+2}}, & y \neq 1,\ 0 \prec y \prec 2, \\ -\delta^3(t), & y = 2, \end{cases} \qquad (6)$$

in which $\Gamma$ denotes the gamma function, and $\delta$ represents the Dirac delta function. We have

$$\delta^n(t) * p(t) = \frac{\partial^n p}{\partial t^n}. \qquad (7)$$

Note that the convolution operation in (5) implies infinite limits on the time integral, and (5) is therefore not causal. To add causality in terms of the time-causal relation (Szabo 1994a), $L_{\alpha,y,t}$ is replaced by

$$S_y(p) = 2H(t)L_{\alpha,y,t}, \qquad 0<y<2, \qquad (8)$$

where $H(t)$ is the Heaviside operator. Assuming that $t=0$ is the initial quiescence instant, we have

$$S_y(p) = \begin{cases} \partial p/\partial t, & y = 0, \\ -\dfrac{2\Gamma(y+2)\cos[(y+1)\pi/2]}{\pi} \displaystyle\int_0^t \dfrac{p(\tau)}{(t-\tau)^{y+2}}\,d\tau, & 0 \prec y \prec 2, \\ -\partial^3 p/\partial t^3, & y = 2. \end{cases} \qquad (9)$$

Note that *y*=1 term in (9) is included in the non-integer expression. We call $S_y(p)$ the Szabo operator. (5) is revised as

$$\Delta p = \frac{1}{c_0^2}\frac{\partial^2 p}{\partial t^2} + \frac{2\alpha_0}{c_0}S_y(p). \qquad (10)$$

It is observed that the Szabo operator encounters the hyper-singular improper integral when *y*≠0,2 and does not explicitly show how to implement the initial conditions.

Szabo (1994a) and Szabo and Wu (2000) also briefly mentioned that the Szabo's model shared some similarity with the fractional derivative equation models of the frequency-dependent attenuation such as the Bagley-Torvik attenuation model (Bagley and Torvik, 1983). One essential difference between both types of anomalous attenuation models, however, is that the Szabo's model guarantees the positive definition operation of non-integer *y* by using (Lighthill, 1962)

$$\begin{aligned}FT^{-}(|\omega|^y) &= \Gamma(y+1)\cos[(y+1)\pi/2]/(\pi|t|^{y+1}) \\ &= s(t),\end{aligned} \qquad (11)$$

where $FT^-$ is the inverse Fourier transform operation. To deal with the odd integer $y$ cases, Szabo (1994a) employed the inverse Fourier transform of the sign function (Lighthill, 1962)

$$FT^-\left(\omega^y \text{sgn}(\omega)\right) = \Gamma(y+1)(-1)^{(y+1)/2}/\left(\pi t^{y+1}\right). \tag{12}$$

We find that (12) can be actually covered within (11). The inverse Fourier transforms (11) and (12) play an essential part in Szabo's creating the causal lossy wave equation model (5).

Compared with the Szabo's models, the fractional derivative attenuation models in the sense of Riemann-Liouville or Caputo fails to recover dispersion equation (3) due to

$$FT^-\left[(-i\omega)^y P\right] = \partial^y p/\partial t^y, \tag{13}$$

when $y$ is a fraction or an odd integer.

On the other hand, the fractional derivative equation models in the Caputo sense, however, have an obvious advantage over the Szabo's model in that they inherently regularize the hyper-singular improper integral and naturally implement the initial conditions (Seredynska and Hanyga, 2000). In the following section II we present a model equation combining the merits of the Szabo's model and the Caputo fractional derivative models via a newly-defined positive time fractional derivative.

## II. MODIFIED SZABO'S WAVE EQUATION MODEL WITH THE CAPUTO FRACTIONAL DERIVATIVE

In terms of the Riemann-Liouville fractional derivative $D_*$, defined in the Appendix, we can rewrite the Szabo's operator (9) as

$$S_y(p) = \begin{cases} \partial p/\partial t, & y = 0, \\ \dfrac{4}{\pi}\int_0^t \dfrac{p(\tau)}{(t-\tau)^3}d\tau, & y = 1, \\ -\dfrac{2\Gamma(y+2)\Gamma[-(y+1)]\cos[(y+1)\pi/2]}{\pi}D_*^{y+1}p, & y \neq 1, \ 0 \prec y \prec 2, \\ -\partial^3 p/\partial t^3, & y = 2. \end{cases} \qquad (14)$$

Note that the initial quiescent instant is set zero in (14). It is found from (14) that the Szabo's model is very much similar to the Riemann-Liouville derivative model for the non-integer power $y$. As discussed in the Appendix, the Caputo fractional derivative is more appropriate for modeling of engineering initial value problems, while the Riemann-Liouville fractional derivative model is not the well-posedness (Seredynska and Hanyga, 2000). Thus, we modify the Szabo's loss operator via the Caputo fractional derivative as

$$Q_y(p) = \begin{cases} \partial p/\partial t, & y = 0, \\ \dfrac{2}{\pi}\int_0^t \dfrac{D^2 p(\tau)}{(t-\tau)}d\tau, & y = 1, \\ -\dfrac{2\Gamma[-(y+1)]\Gamma(y+2)\cos[(y+1)\pi/2]}{\pi}D^{y+1}p, & y \neq 1, \ 0 \prec y \prec 2, \\ -\partial^3 p/\partial t^3, & y = 2. \end{cases} \qquad (15)$$

Note that when 0<$y$<2, the Riemann-Liouville derivative in (14) is replaced by the Caputo derivative in (15), and when $y=1$, the definition $Q_y(p)$ still has the singularity at the origin. Substituting the loss operator $Q_y(p)$ instead of $S_y(p)$ into (10), we have the linear wave equation model of the power law frequency-dependent attenuation

$$\Delta p = \frac{1}{c_0^2}\frac{\partial^2 p}{\partial t^2} + \frac{2\alpha_0}{c_0}Q_y(p). \qquad (16)$$

The model equation (16) eases the hyper-singular improper integral in the Szabo's model (10) through a regularization process of the Caputo derivative (A5 in the Appendix) by invoking the initial conditions naturally. It is worth mentioning that (16) has the same dispersion equation as the Szabo's model equation (10) since Eq. (A11) in the Appendix shows that the Caputo fractional derivative differs from the Riemann-Lioville fractional derivative only in that the former augments a polynomial series with the initial conditions as coefficients, which has no affect on the dispersion equation. Thus, the modified Szabo's model holds the causality of the original Szabo's model.

## III. A POSITIVE TIME FRACTIONAL DERIVATIVE

The positivity of the attenuation (damping) operation required by the decay of the global energy (Matignon et al, 1998) can well be preserved via a positive fractional calculus operator. The space fractional Laplacian (Hanyga, 2001; Chen and Holm, 2002) is the positive definition operator and thus very suitable to describe the anomalous attenuation behaviors, while the traditional time fractional derivative lacks this crucial positive property. In this section, we introduce a positive time fractional derivative in terms of (11), whose Fourier transform characterizes the positive operation as follows

$$FT^+\left(D^{|\eta|}p\right)=|\omega|^\eta P(\omega), \qquad 0<\eta<2, \tag{17}$$

where $FT^+$ is the Fourier transform operation; $P(\omega)$ is the Fourier transform of the pressure signal $p(t)$; and

$$D_*^{|\eta|}p = s(t)*p(t) = \frac{1}{q(\eta)}\int_0^t \frac{p(\tau)}{(t-\tau)^{\eta+1}}d\tau, \tag{18}$$

$$D^{|\eta|}p = \begin{cases} \dfrac{-1}{\eta q(\eta)}\int_0^t \dfrac{D^1 p(\tau)}{(t-\tau)^\eta}d\tau, & 0 \prec \eta \leq 1, \\ \dfrac{1}{\eta(\eta-1)q(\eta)}\int_0^t \dfrac{D^2 p(\tau)}{(t-\tau)^{\eta-1}}d\tau, & 1 \prec \eta \prec 2, \end{cases} \tag{19}$$

where $s(t)$ is defined in (11), and

$$q(\eta) = \frac{\pi}{2\Gamma(\eta+1)\cos[(\eta+1)\pi/2]}. \tag{20}$$

The definitions (18) and (19) of the positive fractional derivative are to combine the Fourier transform relationship (11) respectively with the Riemann-Liouville fractional derivative (A7) and with the Caputo fractional derivative (A8) in the Appendix to hold the Fourier transform relationship (17). (19) can be expanded by the integration by parts into (18) augmented a regularization series. The positive fractional derivative (19) can also further be generalized by

$$D^{|\eta|}u = \begin{cases} \dfrac{-1}{(\eta-2k)q(\eta)}\int_0^t \dfrac{D^{2k+1}u(\tau)}{(t-\tau)^{\eta-2k}}d\tau, & 2k \prec \eta \leq 2k+1, \\ \dfrac{1}{(\eta-2k)[\eta-(2k+1)]q(\eta)}\int_0^t \dfrac{D^{2k+2}u(\tau)}{(t-\tau)^{\eta-(2k+1)}}d\tau, & 2k+1 \prec \eta \prec 2k+2, \end{cases} \quad (21)$$

where $k$ is a non-negative integer. We also have

$$D^{|\eta|+q}p = D^{|\eta|}D^q p \neq D^q D^{|\eta|}p, \quad (22)$$

$$D_*^{|\eta|+q}p = D^q D_*^{|\eta|}p \neq D_*^{|\eta|}D^q p, \quad (23)$$

where $q$ is a positive integer number. Note that $D^{|1|}p \neq D_*^{|1|}p \neq D^1 p$. It is straightforward to have

$$FT^+\left(D^{|\eta|+q}p\right) = (-i\omega)^q |\omega|^\eta P(\omega), \quad 0<y<2. \quad (24)$$

In terms of the new fractional derivative definition, the loss operator $Q_y(p)$ from (15) is rewritten as

$$Q_y(p) = \begin{cases} D^1 p, & y = 0, \\ D^{|y|+1} p, & 0 \prec y \prec 2, \\ -D^3 p, & y = 2. \end{cases} \quad (25)$$

where *y*=0 corresponds to the damped wave equation, also known as the electromagnetic equation (Szabo, 1994a).; and *y*=2 leads to the thermoviscous equation (Backstock, 1967; Pierce, 1989), also called the augmented wave equation (Johnson and Dudgeon, 1993). 0<y<2 occurs in most cases of practical interest. It is noted that the fractional derivative Burgers equation for anomalous attenuations presented by Ochmann and Makarov (1993) requires a sign change of viscous coefficient between 0≤*y*<1 and 1<*y*≤2, whereas the definition (19) of the positive time fractional derivative avoids the sign change across *y*=1 in (25).

If we define

$$D^{|\eta|} p = \begin{cases} p, & \eta = 0, \\ -D^2 p, & \eta = 2, \end{cases} \quad (26)$$

then the modified Szabo equation (16) can be rewritten as

$$\Delta p = \frac{1}{c_0^2} \frac{\partial^2 p}{\partial t^2} + \frac{2\alpha_0}{c_0} D^{|y|+1} p. \quad (27)$$

Note that the definition (19) implies that unlike the even order derivative, the positive time fractional derivative of the odd order is not equivalent to the corresponding standard derivative.

## IV. CONCLUSIONS

The modified Szabo's model equation (16) includes the Caputo fractional derivative, the existence, uniqueness, and well-posedness (i.e. continuous dependence on data) of whose type have been well proved (Seredynska and Hanyga, 2000). The corresponding numerical solution can also be carried out with the well-developed standard numerical methods for the Caputo fractional derivative equations (Diethelm, 1997, 2000). It is worth stressing that the modified Szabo's model is closely related to the singular Votlerra equation of the second kind, of which the numerical solutions are discussed in Press et al (1992) and many other publications.

In this study, only the linear model of wave equation was involved. It is straightforward to extend the present strategy to a variety of the corresponding parabolic equations and nonlinear equation models developed by Szabo (1993) and (1994a). It is also noted that the mathematical models for anomalous dissipation behaviors are mostly phenomenological (Adhikari, 2000). In other words, these models are to describe attenuation phenomena but do not necessarily reflect various physical and chemical mechanisms behind scene. Broadly speaking, there are two types of attenuation (damping) model methodology, space and time operations. The present time-domain model underlies the memory effect, also often called heredity, relaxation or hysteresis in some publications, which depends on the past history of motion. On the other hand, the space fractional Laplacian models reflect the fractal microstructures of media and can describe the frequency power law attenuation quite well (Hanyga, 2001; Chen and Holm, 2002). When $y \geq 1$, the modified Szabo's time-domain models require the second order initial condition, which is not available in many practical cases, and it is thus reasonable to use the space-domain attenuation model instead. When $y<1$, the present modified Szabo's model is the model of choice since its numerical solution is relatively easier than the fractional Laplacian space-domain model.

**ACKNOWLEDGMENTS**

**APPENDIX**

The Riemann-Liouville fractional integral is an essential concept to understand the fractional derivative and is given by (Samko et al, 1987; Diethelm, 2000)

$$J^q\{\psi(t)\} = \frac{1}{\Gamma(q)} \int_a^t \frac{\psi(\tau)}{(t-\tau)^{1-q}} d\tau, \qquad 0<q, \qquad (A1)$$

where $q$ and $a$ are real valued. The corresponding Riemann-Liouville fractional derivative is interpreted as

$$D_*^\lambda\{\psi(t)\} = \frac{d}{dt}\left[J^{1-\lambda}\{\psi(t)\}\right], \qquad 0<\lambda<1, \qquad (A2)$$

which can be further elaborated into

$$D_*^\lambda \{\psi(t)\} = \frac{1}{\Gamma(1-\lambda)} \frac{d}{dt} \int_a^t \frac{\psi(\tau)}{(t-\tau)^\lambda} d\tau$$
$$= \frac{1}{\Gamma(-\lambda)} \int_a^t \frac{\psi(\tau)}{(t-\tau)^{1+\lambda}} d\tau \qquad (A3)$$
$$= J^{-\lambda} \{\psi(t)\}.$$

The Riemann-Liouville fractional derivative, however, has some notable disadvantages in engineering applications such as the hyper-singular improper integral, where the order of singularity is higher than the dimension, and nonzero of the fractional derivative of constants, e.g., $D^\lambda 1 \neq 0$, which would entail that dissipation does not vanish for a system in equilibrium (Samko et al, 1987; Seredynska and Hanyga, 2000) and invalidates the causality. The Caputo fractional derivative has instead been developed to overcome these drawbacks (Caputo, 1967; Caputo and Mainardi, 1971) as defined below

$$D^\lambda \{\psi(x)\} = J^{1-\lambda} \left[ \frac{d}{dx} \psi(x) \right]. \qquad (A4)$$

Integration by part of (A4) yields

$$D^\lambda \{\psi(t)\} = \frac{1}{\Gamma(1-\lambda)} \int_a^t \frac{1}{(t-\tau)^\lambda} \frac{d\psi(\tau)}{d\tau} d\tau$$
$$= D_*^\lambda \{\psi(t)\} - \frac{\psi(a)}{\Gamma(1-\lambda)} \frac{1}{(t-a)^\lambda}. \qquad (A5)$$

It is observed that the right-hand second term in (A5) regularizes the fractional derivative to avoid the potential divergence from singular integration at $t=a$. In addition, the Caputo fractional differentiation of a constant results in zero. For instance, Seredynska and Hanyga (2000) pointed out that the solution of the following time fractional derivative model

$$D^2 u + \gamma D^{1+\eta} u + F(u) = 0, \quad 0 < \eta \leq 2 \tag{A6}$$

is not $C^2$ smooth if the Caputo fractional derivative is replaced by the Riemann-Liouville fractional derivative. The Caputo fractional derivative also implicitly includes the initial function value at $t=a$ as shown in (A5), which is convenient in handling initial value problem. Therefore, the fractional derivative defined in the Caputo sense is essential in the modeling of various anomalous attenuation behaviors (Makris and Constantinou, 1991; Seredynska and Hanyga, 2000).

The fractional derivatives in the Riemann-Liouville and Caputo senses can respectively in general be expressed as

$$D_*^\mu \{\psi(x)\} = D^m \left[ J^{m-\mu} \psi(x) \right], \tag{A7}$$

$$D^\mu \{\psi(x)\} = J^{m-\mu} \left[ D^m \psi(x) \right]. \quad m-1 \prec \mu \leq m, \tag{A8}$$

where $m$ is an integer, and $D^m \psi = d^m \psi / dt^m$. The Riemann-Liouville fractional derivative (A7) can be recast as (Seredynska and Hanyga, 2000)

$$D_*^\mu \psi = \theta_{-\mu-1} * \psi, \tag{A9}$$

where

$$\theta_\mu(t) = |t|^\mu / \Gamma(\mu+1). \tag{A10}$$

By integration by parts, the Caputo fractional derivative (A8) can be reduced to

$$\begin{aligned} D^\mu \psi &= \theta_{-\mu-1} * \psi - \sum_{k=0}^{m-1} D^k \psi(0) \theta_{k-\mu} \\ &= D_*^\mu \psi - \sum_{k=0}^{m-1} D^k \psi(0) \theta_{k-\mu}, \end{aligned} \tag{A11}$$

where the first term of the right hand side is in fact the Riemann-Liouville fractional derivative.